\documentclass[aps,prd]{revtex4}

\usepackage{graphicx}
\let\l=\left
\let\r=\right
\def\be{\begin{equation}}
\def\ee{\end{equation}}
\def\bea{\begin{eqnarray}}
\def\eea{\end{eqnarray}}

\begin{document}

\title{Light rings and light points of boson stars}

\author{Philippe Grandcl\'ement}
\affiliation{LUTH, Observatoire de Paris, PSL Research University, CNRS, Universit\'e Paris Diderot, 5 place Jules Janssen, 92190 Meudon Cedex, France.}


\begin{abstract}

This work is devoted to the study of photon trajectories around rotating boson stars. The basic properties of boson star models are given with a particular emphasis on the high compactness those objects can have. Using an effective potential method, circular orbits of photons around rotating boson stars are then obtained, at least for relativistic enough configurations. A particular class of light rings, in which the photons are at rest on a stable orbit, is exhibited. 
By this one means that the associated worldline is collinear to the Killing vector corresponding to the asymptotic time translation symmetry. It is proposed to call those orbits {\em light points}. Their existence is very specific to boson stars and the link between light points and ergoregions is investigated.

\end{abstract}

\maketitle

\section{Introduction}

Boson stars were first introduced in the late 1960s in \cite{BonazP66,Kaup68,RuffiB69}. They consists of a complex scalar field coupled to gravity. Various families of boson stars have been constructed, with various potentials for the scalar field, with or without rotation, in various dimensions (see Refs. \cite{Jetze92,LeeP92,SchunM03,LieblP12} and references therein)...

One of the main motivations for the study of boson stars is the fact that they can act as black hole mimickers \cite{GuzmaR09}. What this means is that they can have a large mass with a small size without developing any horizon or singularity. Moreover, boson stars have no hard surface, the scalar field interacting with ordinary matter only through gravitation. Possible scenarios for the formation of boson stars have been studied. There has also been a number of studies that proposed observational means of discriminating boson stars from black holes. This could, for instance, be done by direct imaging of the Galactic center \cite{VinceMGGS16}. Observations of gravitational waves could also be a fruitful way to investigate the nature of the compact objects observed in the Universe \cite{KesdeGK04, MacedPCC13}.

Apart from being an alternative to black holes, boson stars are very interesting test beds for a lot of situations in which intense gravitational fields play a role. For instance, a class of peculiar orbits for massive particles is investigated in Ref. \cite{GrandSG14}. Other specific effects concerning the imaging of accretion disks around boson stars are studied in Refs. \cite{MeliaVGGMS15, VinceMGGS16}.

The main goal of the current paper is to investigate the existence of light rings around boson stars. Those are closed (circular) orbits of photons. They can only appear for very compact objects (dubbed ultracompact in Ref. \cite{CardoCMOP14}). It is known that black holes admit light rings and that they lie outside the horizon. The link between light rings and various instabilities was explored in a recent study by V. Cardoso {\em et al.} \cite{CardoCMOP14}. Orbits of photons around boson stars or black holes with scalar hair have already been discussed in Refs. \cite{CunhaHRR15, CunhaGHRRW16}, especially in the context of imaging those objects. In this work results are expanded to boson stars with higher angular momentum.

In a recent paper \cite{CardoFP16}, the connection between light rings and the gravitational waves observed by the LIGO observatory \cite{ligo} is studied. The first oscillations of the ringdown signal are shown to be the signature of the presence of a light ring more than of an apparent horizon. Two objects with exactly the same structure in terms of light rings would generate similar early ringdown signal, regardless of the presence of a horizon.

This paper is organized as follows. In Sec. \ref{s:models} the reader is reminded about models of boson stars. In particular, the high compactness of such objects is exhibited. Section  \ref{s:rings} describes the study of light rings by means of an effective potential method. In Sec. \ref{s:points}, a particular class of light rings, in which the photons are {\em at rest} and on a stable orbit, is investigated. Those {\em light points} are a new class of orbits, typical of boson stars. Indeed, their appearance requires an intense gravitational field but also no horizon. In the case of black holes, stationary orbits are located exactly on the horizon and are unstable. Conclusions are given in Sec. \ref{s:ccl}.

\section{Boson star models} \label{s:models}

In this section we recall the setting used in Ref. \cite{GrandSG14} to compute boson star spacetimes. Boson stars consist of a complex scalar field $\Phi$ coupled to gravity. This is achieved by considering the action 
\be\label{e:action}
S = \int \l({\mathcal L}_g + {\mathcal L}_\Phi\r) \sqrt{-g} \rm{d}x^4,
\ee
where ${\mathcal L}_g$ is the Hilbert-Einstein Lagrangian and ${\mathcal L}_\Phi$ is the Lagrangian of the complex scalar field. They are given by the standard expressions

\bea
 {\mathcal L}_g &=& \frac{1}{16\pi}R \\
{\mathcal L}_\Phi &=& -\frac{1}{2}\l[\nabla_\mu \Phi \nabla^\mu \bar{\Phi} + V\l(\l|\Phi\r|^2\r)\r].
\eea

$\nabla$ denotes the covariant derivative associated to ${\bf g}$, and $R$ is the Ricci scalar. $V\l(\l|\Phi\r|^2\r)$ is a potential that can be chosen to construct various types of boson stars. In this paper one considers the simplest possible choice in which the scalar field is a free field, which implies that

\be
V\l(\l|\Phi\r|^2\r) = \frac{m^2}{\hbar^2} \l|\Phi\r|^2.
\ee

$m$ is the mass of the individual boson and the factor $m/\hbar$ then appears as a scale factor for the various quantities (distances, masses, etc). Throughout this paper geometric units are used, such that $G=1$ and $c=1$.

Rotating boson stars are computed by demanding that the scalar field takes the form
\be
\Phi = \phi\l(r,\theta\r) \exp\l[i\l(\omega t-k\varphi\r)\r],
\ee
where $\phi$ is the amplitude of the field (and so real), $\omega$ is a real constant (smaller than $m/\hbar$) and $k$ an integer known as the rotational quantum number. Because of the $U\l(1\r)$ symmetry of the action (\ref{e:action}), it does not depend on $t$ and $\varphi$, leading to axisymmetric solutions. This is why the amplitude $\phi$ depends on $\l(r,\theta\r)$ only. The numbers $\omega$ and $k$ appear as parameters of the various boson stars. Let us point out that the case $k=0$ corresponds to spherically symmetric boson stars.

For the metric, quasi-isotropic coordinates are well adapted to the symmetry of the problem. The metric reads
\be
g_{\mu\nu} {\rm d}x^\mu {\rm d}x^\nu = -N^2 {\rm d}t^2 + A^2 \l({\rm d}r^2 + r^2{\rm d}\theta^2\r) + B^2 r^2\sin^2\theta\l({\rm d}\varphi+ \beta^\varphi{\rm d}t\r)^2.
\ee

In the 3+1 language, $N$ is the lapse, the shift vector is $\beta^i= \l(0,0,\beta^\varphi\r)$ and the spatial metric is $\gamma_{ij} = {\rm diag}\l(A^2, B^2r^2, B^2r^2\sin^2\theta\r)$. The unknowns are the functions $N$, $A$, $B$ and $\beta^\varphi$, all of them depending only on $\l(r,\theta\r)$. The unknown fields obey the Einstein-Klein-Gordon system which results from the variation of the action (\ref{e:action}) with respect to both the metric ${\bf g}$ and the scalar field $\Phi$. 

\begin{figure}[!hbtp]
\includegraphics[width=12cm]{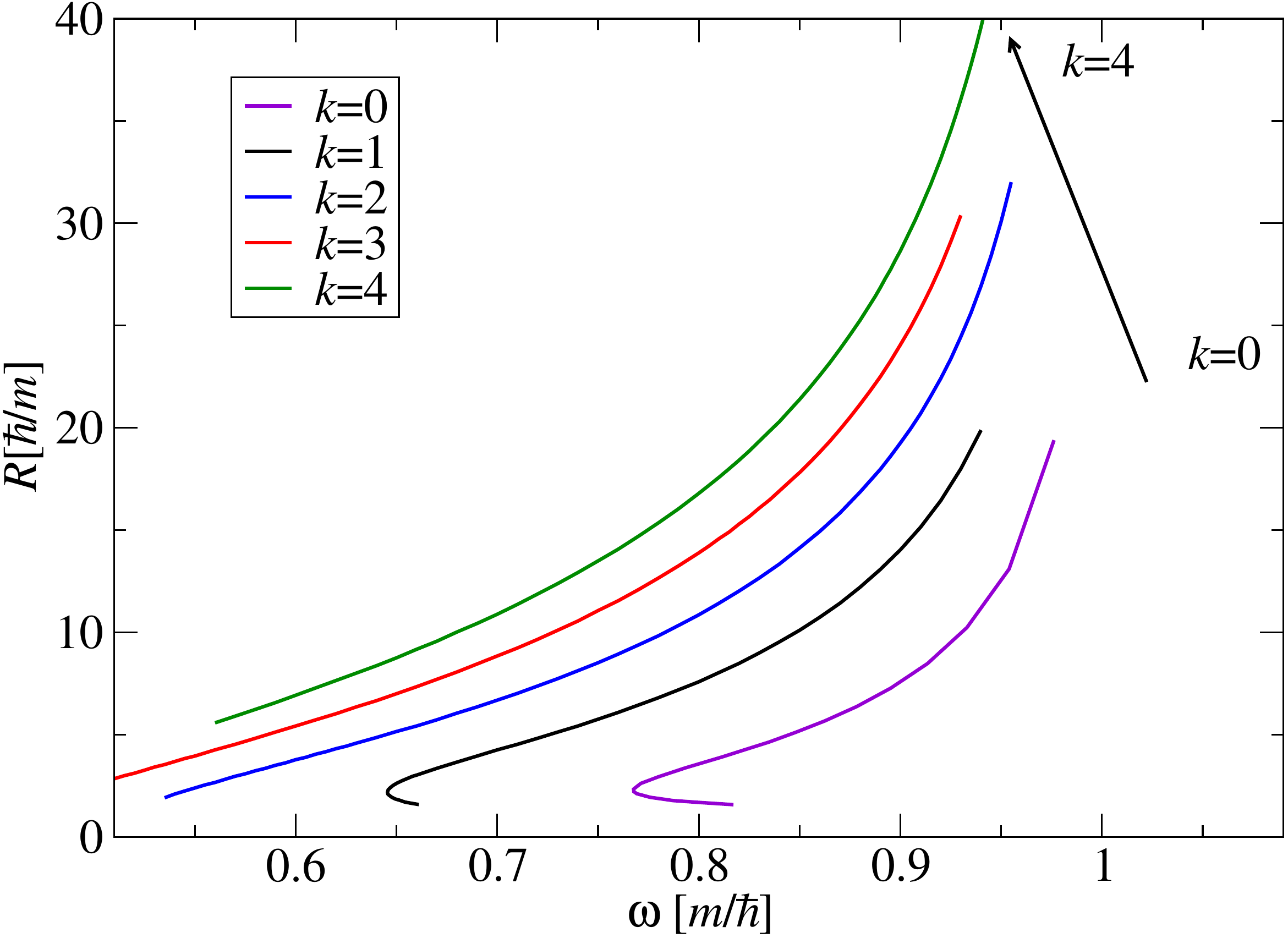}
\caption{\label{f:rad} Radii of the various boson stars, defined as the radius for which the field is $0.1$ times its maximal value, in the equatorial plane.}
\end{figure}

Those equations are solved by means of highly accurate spectral methods, implemented by the KADATH library \cite{kadath}. This enables us to compute sequences of rotating boson stars for $k$ ranging from $0$ to $4$. Numerical methods and tests are given in detail in Ref. \cite{GrandSG14}. We present here the sequences in a different way: by plotting the compactness of the various configurations. To do so, one needs to define the radius of the boson stars. This cannot be done unambiguously because the scalar field extends up to infinity and those objects have no true surface. In the following, the radius $R$ of the boson star is defined as the radius for which the field is $0.1$ times its maximal value, in the equatorial plane. The precise value of the thus-defined radius obviously depends on the value of the threshold (here $0.1$). However, because of the fast decay of the field far from the origin, it is a small effect. Figure \ref{f:rad} shows the radii of the various boson star models. One can clearly see the known fact that the size of a boson star increases with $\omega$ and $k$.

\begin{figure}[!hbtp]
\includegraphics[width=12cm]{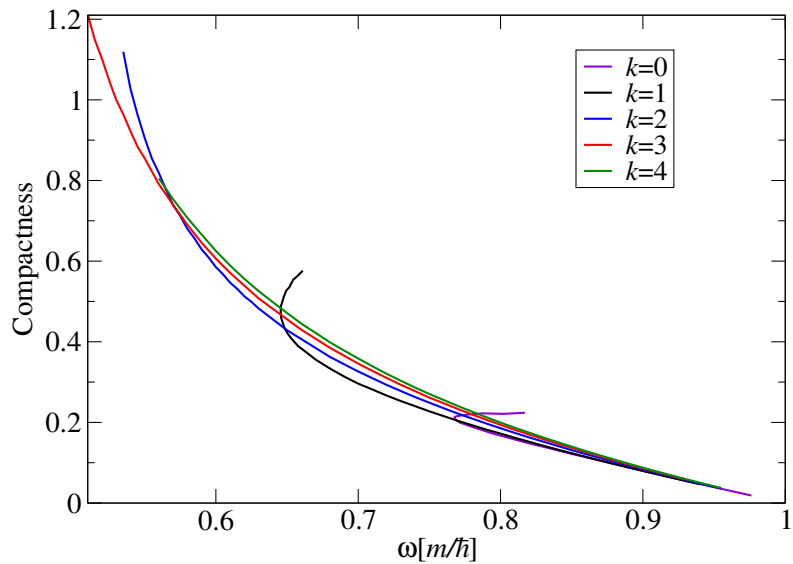}
\caption{\label{f:comp} Compactness of the various boson stars.}
\end{figure}

With this definition of the radius, it is possible to compute the compactness which is the adimensional ratio $M/R$, where $M$ is the standard Arnowitt-Deser-Misner mass. Compactness is shown in Fig. \ref{f:comp}. First one can notice that the result depends very moderately on the value of $k$. The variations of both the mass and radius with $k$ are such that the compactness depends almost only on $\omega$. One can also notice that it can reach very high values of the order unity (remember that typical neutron stars have a compactness of about 0.2), illustrating the fact that boson stars are indeed good black hole mimickers. Let us also mention that spherically symmetric boson stars (i.e. for which $k=0$) can only reach compactness of about $0.25$. This is mainly due to the fact that they cannot attain small enough values of $\omega$, due to the turning point in the sequences (see Fig. \ref{f:rad} for instance). The results obtained in the case $k=1$ are consistent with those shown in Fig. 2 of Ref. \cite{HerdeR15}.

\section{Light rings} \label{s:rings}

Given the high compactness of boson stars, they can be expected to have light rings which are closed orbits of photons. In Ref. \cite{CunhaHRR15} light rings are mentioned (see Fig. 1 in that reference) for both $k=0$ and $k=1$ boson stars, in the context of imaging them. We extend the study of light rings to boson stars with higher $k$. Mathematically speaking one needs to find closed null geodesics of the spacetime. Given the fact that boson stars are axisymmetric objects, one looks for circular orbits in the equatorial plane, as in the case of a Kerr black hole. In this work such orbits are localized using an effective potential method. Let us mention that, in Ref. \cite{CunhaGHRRW16} similar techniques are employed to study photon orbits, in a somewhat more general setting (i.e. for orbits not confined in the orbital plane). Their results appear to be consistent with this work, in the case $k=1$.

The effective potential method is a standard technique to study orbits around axisymmetric and stationary objects and it proceeds as follows. Let us call $U^\mu$ the tangent vector of a photon trajectory. As the orbits are in the equatorial plane, it reduces to $\l(U^t, U^r, 0, U^\varphi\r)$. Boson star spacetimes admit two independent Killing vectors : $\l(\partial_t\r)^\mu$ and $\l(\partial_\varphi\r)^\mu$. Their existence leads to two conserved quantities being the scalar products of $U^\mu$ with the two Killing vectors:

\bea
U_\mu \l(\partial_t\r)^\mu &=& \l(-N^2+ B^2 r^2 \beta^{\varphi \, 2}\r) U^t + B^2 r^2 \beta^\varphi U^\varphi = -E \\
U_\mu \l(\partial_\varphi\r)^\mu &=& B^2 r^2 \beta^\varphi U^t + B^2 r^2 U^\varphi = L,
\eea
where $E$ and $L$ are two constants along the geodesics. From those equations, one can express the components $U^t$ and $U^\varphi$, as functions of $E$ and $L$ :
\bea
\label{e:Ut}
U^t &=& \frac{\beta^\varphi L + E}{N^2} \\
\label{e:Up}
U^\varphi &=& \frac{L}{B^2 r^2} - \frac{\beta^\varphi}{N^2}\l(\beta^\varphi L +E\r).
\eea
Injecting those expressions into the fact that the geodesic is null (i.e. $U_\mu U^\mu=0$) leads to an equation on $U^r$. It can be put into the form $\l(U^r\r)^2 + V_{\rm eff} = 0$ where $V_{\rm eff}$ is an effective potential, given by 

\be
\label{e:veff}
V_{\rm eff} = \frac{1}{A^2} \l[-\frac{\l(\beta^\varphi L +E\r)^2}{N^2} + \frac{L^2}{B^2 r^2}\r].
\ee

Circular orbits are such that $V_{\rm eff}=0$ and $\partial_r V_{\rm eff}=0$. The second condition is necessary to prevent being only at the periastron or the apoastron of an elliptic orbit. The first condition leads to
\be
\label{e:EsL}
\frac{E}{L} = -\beta^\varphi + \epsilon \frac{N}{Br},
\ee
where $\epsilon = \pm 1$. Contrary to the massive particle case, $E$ and $L$ are not constrained independently but only via their ratio. This is basically due to the fact that there is no scale of mass in the case of photons. 

The condition $\partial_r V_{\rm eff} = 0$, along with $V_{\rm eff}=0$ gives an equation involving the radius of the orbit and the various metric fields :
\be
I\l(r\r) = \l(\epsilon \frac{\partial_r \beta^\varphi}{NB}\r) r^2 + \l(\frac{\partial_rB}{B^3}-\frac{\partial_rN}{NB^2}\r) r + \frac{1}{B^2} = 0.
\ee

\begin{figure}[!hbtp]
\includegraphics[width=12cm]{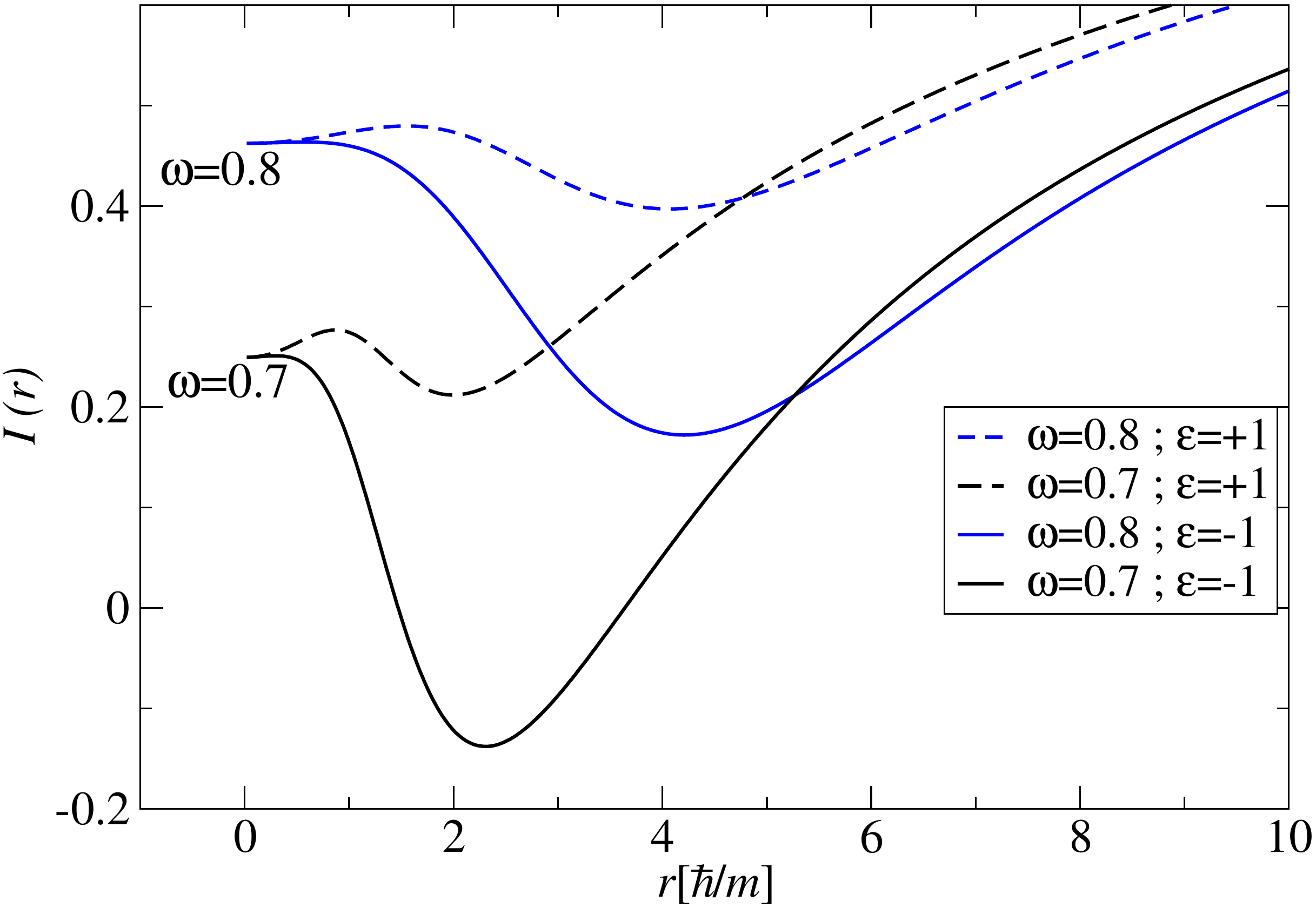}
\caption{\label{f:func} Function $I\l(r\r)$ for $k=1$. The blue curves correspond to $\omega=0.8$ and the black ones correspond to $\omega=0.7$. The dashed lines denote the cases in which $\epsilon=+1$ and the solid lines denote the cases in which $\epsilon=-1$. }
\end{figure}

This is a purely radial equation (remember one works in the equatorial plane), but it is not a second-order equation on $r$, the various fields depending on the radius. Figure \ref{f:func} shows the functions $I\l(r\r)$ for two values of $\omega$, in the case $k=1$. Let us first mention that choosing $\epsilon=+1$ leads to functions $I\l(r\r)$ that always remain strictly positive. This corresponds to the dashed curves of Fig. \ref{f:func}. When $\epsilon=-1$ the functions $I\l(r\r)$ can sometimes vanish, depending on the boson star considered. $I\l(r\r)$ vanishes for the most relativistic boson stars being the ones with small values of $\omega$. It follows that those boson stars admit light rings. In Fig. \ref{f:func}, the boson star $k=1 ; \omega=0.8$ has no light ring (the solid blue curve is always positive), whereas the boson star $k=1 ; \omega=0.7$ has two light rings, corresponding to the two values of $r$ at which $I\l(r\r)$ vanishes (see the solid black curve).

\begin{figure}[!hbtp]
\includegraphics[width=12cm]{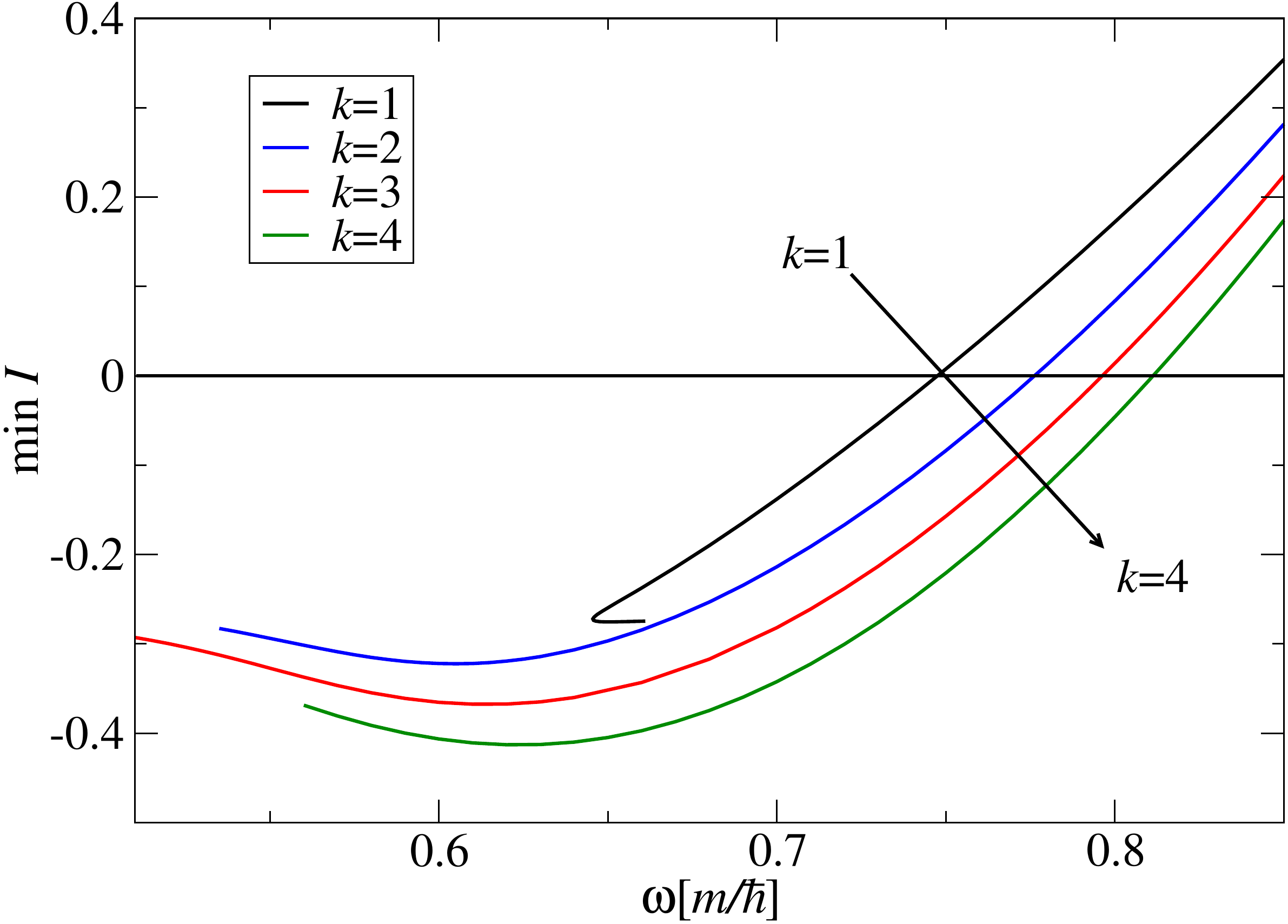}
\caption{\label{f:min} Minimum of the function $I\l(r\r)$, when $\epsilon=-1$. The configurations for which the minimum is below zero admit light rings. }
\end{figure}

Figure \ref{f:min} shows the value of the minimum of $I\l(r\r)$, in the case $\epsilon=-1$. When this minimum is below zero the boson star admits light rings. This corresponds to the configurations with low values of $\omega$ and it can happen for all values of $k>0$. The case $k=0$ is not shown in Fig. \ref{f:min} because the minimum is far above zero. The spherically symmetric boson stars considered in this paper have no light ring. This may be linked to the fact that they cannot reach a high enough compactness. On the other hand, the authors of  Ref. \cite{CunhaHRR15} mentioned the existence of such orbits, even for $k=0$ boson stars. However, this is only possible for objects that are farther along the sequence than the configurations explored here (see Fig. 1 of Ref. \cite{CunhaHRR15}).

\begin{figure}[!hbtp]
\includegraphics[width=12cm]{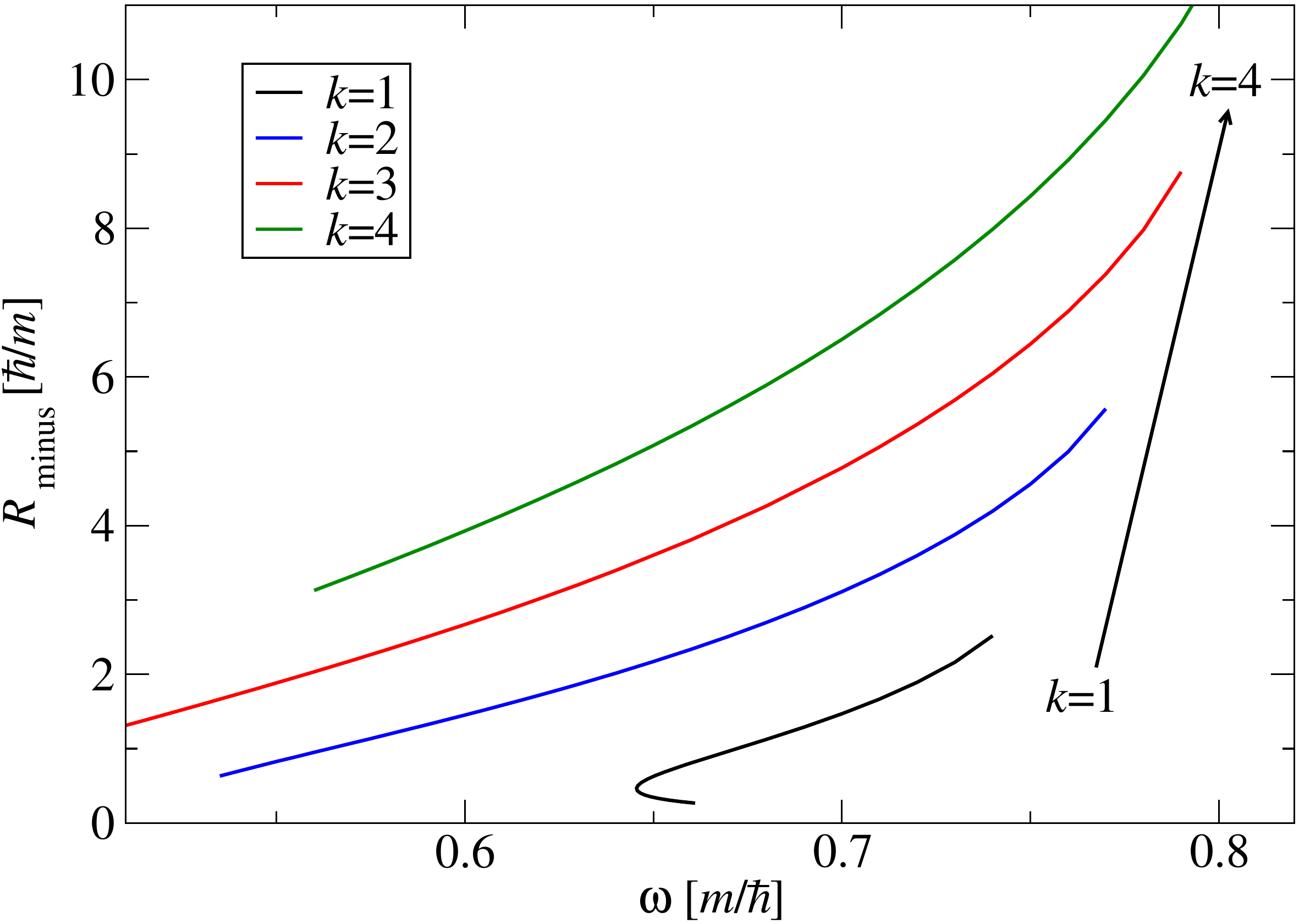}
\includegraphics[width=12cm]{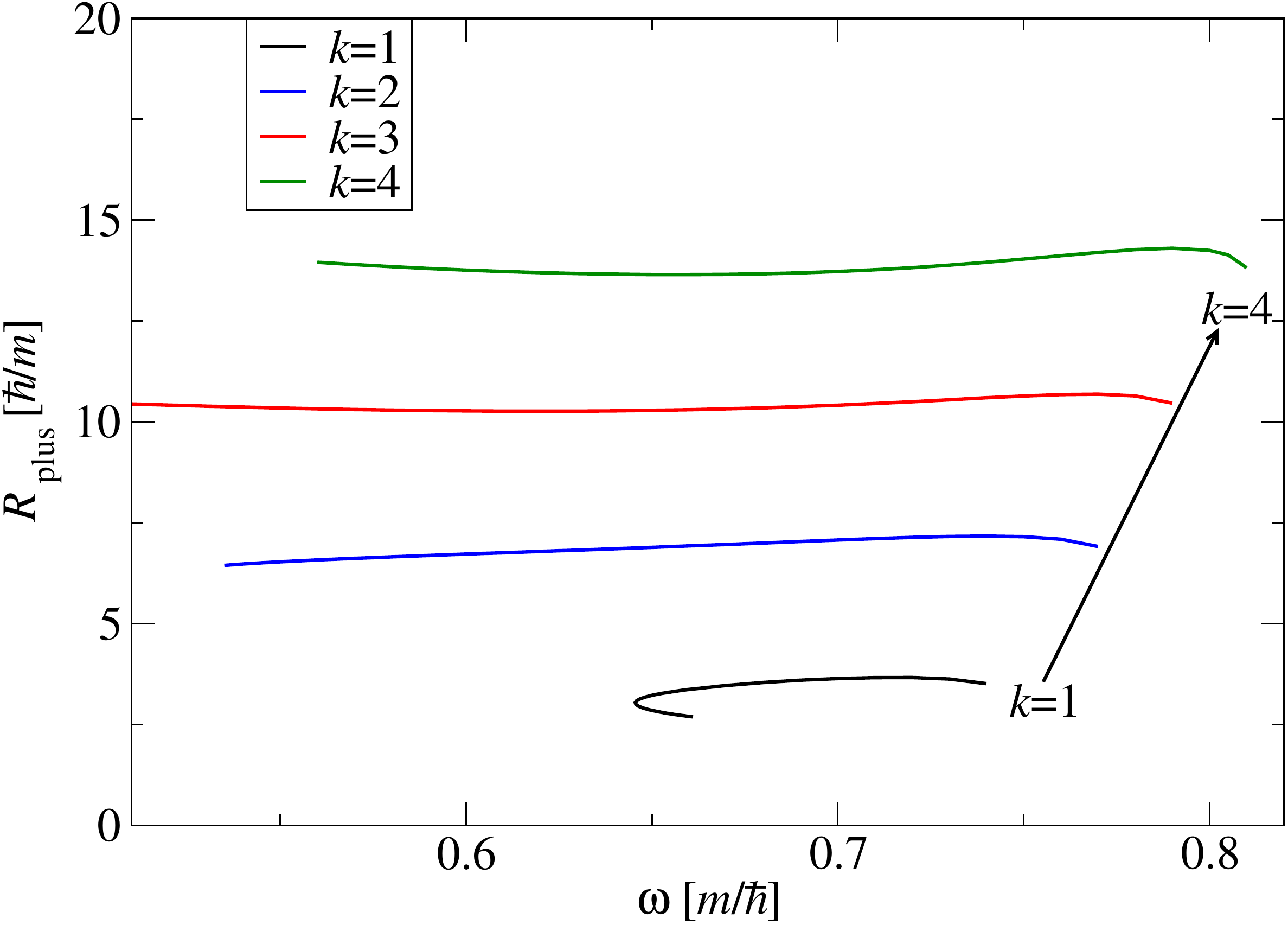}
\caption{\label{f:radlight} Radii of the inner light rings (first panel) and the outer ones (second panel). }
\end{figure}

The shape of the function $I\l(r\r)$ depicted in Fig. \ref{f:func} is very generic : when the minimum of $I\l(r\r)$ is negative there are two light rings, corresponding to two radii $R_{\rm minus}$ and $R_{\rm plus}$. Moreover, it is easy to see that $\partial_r V_{\rm eff}$ and $I\l(r\r)$ have opposite signs. It follows that $R_{\rm minus}$ corresponds to a minimum of $V_{\rm eff}$ and is stable whereas $R_{\rm plus}$ corresponds to a maximum of $V_{\rm eff}$ and so to an unstable orbit. The radii of the light rings are shown in Fig. \ref{f:radlight}.

\begin{figure}[!hbtp]
\includegraphics[width=12cm]{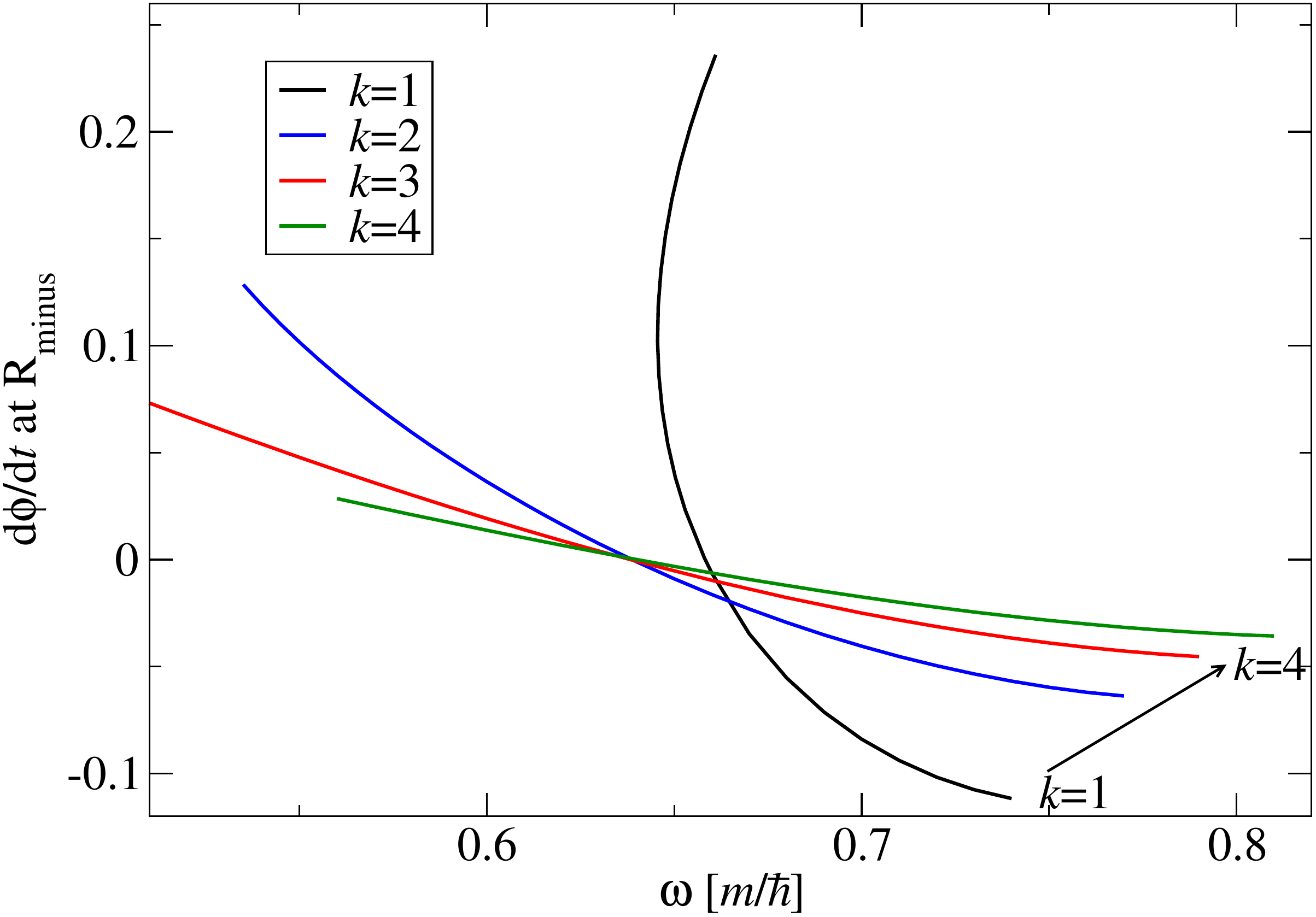}
\includegraphics[width=12cm]{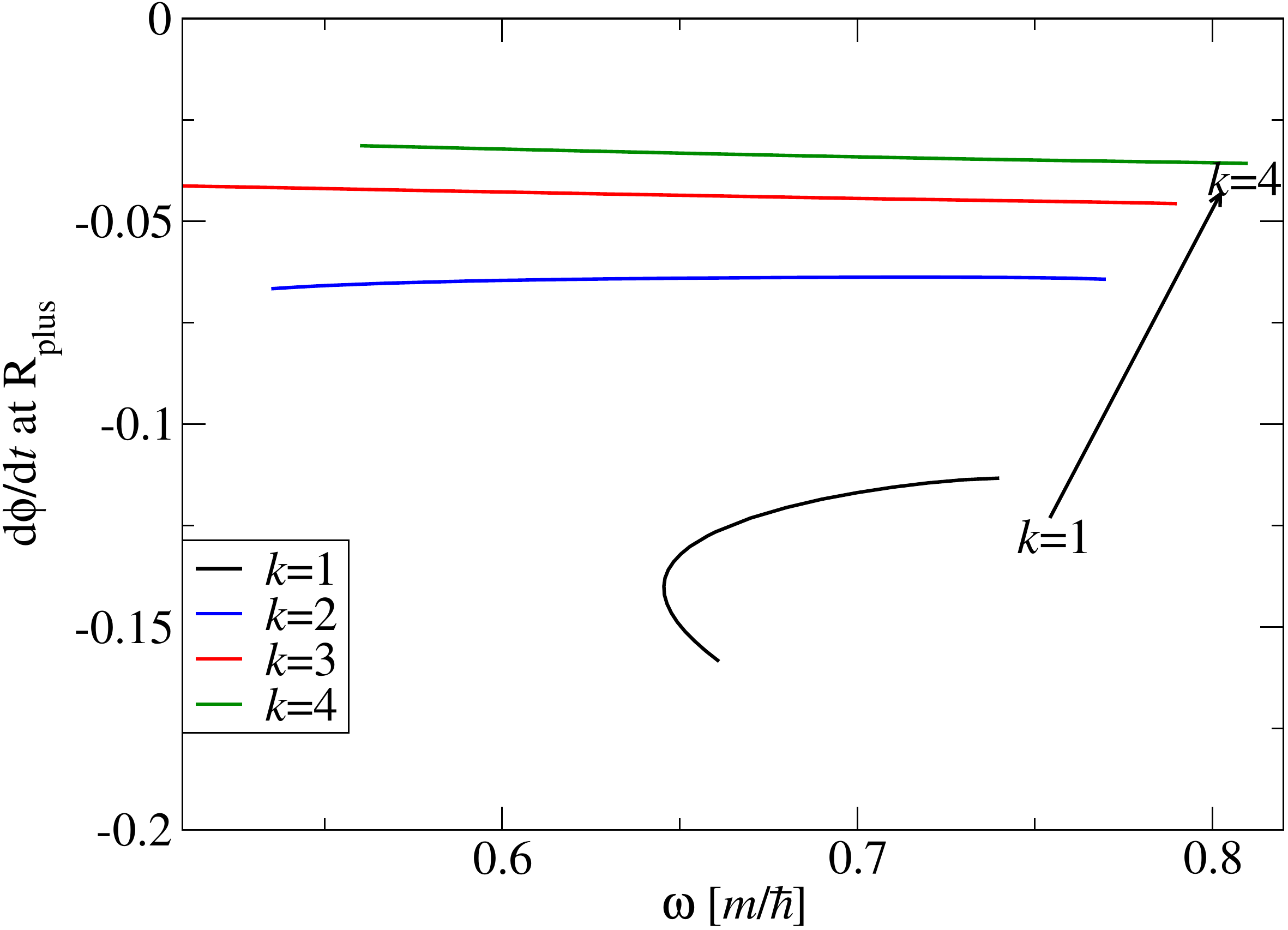}
\caption{\label{f:dpdt} Orbital frequency $\displaystyle\frac{{\rm d}\varphi}{{\rm d}t}$ of the inner light rings (first panel) and the outer ones (second panel). }
\end{figure}

By inserting Eq. (\ref{e:EsL}) into Eqs. (\ref{e:Ut}) and (\ref{e:Up}) one can find the value of $\displaystyle\frac{{\rm d}\varphi}{{\rm d}t}$ for the light rings. It leads to
\be
\label{e:dpdt}
\frac{{\rm d}\varphi}{{\rm d}t} = -\frac{N}{Br} - \beta^\varphi.
\ee
The right-hand side of Eq. (\ref{e:dpdt}) must be evaluated at $R_{\rm minus}$ or $R_{\rm plus}$ depending on the light ring considered. The tangent vector of the trajectories is then given by $U^\mu = \l(1, 0, 0, \displaystyle\frac{{\rm d}\varphi}{{\rm d}t}\r)$ where the parameter along the trajectory has been chosen to be the coordinate $t$. Orbital frequencies of the inner (respectively outer) light rings are shown in Fig. \ref{f:dpdt}. They are discussed further in Sec. \ref{s:points}.

\section{Light points}\label{s:points}

The most striking result from Fig. \ref{f:dpdt} is the fact that the orbital frequency of the inner light ring can change sign. This implies that it can even vanish, for certain boson stars. The associated trajectories are simply given by $U^\mu = \l(1, 0, 0, 0\r)$. In term of spatial coordinates $\l(r,\theta,\varphi\r)$, those photons are not moving. They correspond more to light points than light rings. This effect is more than just a coordinate effect. Indeed those photons can be said to be at rest in the sense that their worldline is collinear to the Killing vector that corresponds to the asymptotic time translation symmetry. This notion of ``at rest'' is not new and is indeed identical to the one used when defining what an ergoregion is.

Even if those light points seem to be peculiar, a detailed study can confirm their existence and location. One first condition is that the vector $U^\mu = \l(1, 0, 0, 0\r)$ must be a null vector. Given the fact that the quasi-isotropic metric is diagonal this condition is $g_{tt}=0$. It is the same as the one defining the boundary of an ergoregion (see Sec. IV-E of Ref. \cite{GrandSG14} for a detailed study of ergoregions of boson stars). So light points must be on the boundary of an ergoregion.

However, not all the null curves are geodesics. The geodesic equation is $\displaystyle\frac{{\rm d}U^\mu}{{\rm d}t} + \Gamma^\mu_{\alpha\beta} U^\alpha U^\beta = C\l(t\r) U^\mu$ where $C\l(t\r)$ is a function along the curve. Note that, {\em a priori}, one cannot make $C\l(t\r)=0$ because there is no guarantee that $t$ is an affine parameter. $\Gamma$ denotes the four-dimensional Christoffel symbols. Inserting $U^\mu = \l(1, 0, 0, 0\r)$ in the geodesic equation leads to the condition

\be
\label{e:geo}
\Gamma^\mu_{tt} = \frac{1}{2} g^{\mu \alpha} \l[-\partial_\alpha g_{tt}\r] = C\l(t\r) U^\mu,
\ee
where all the time derivatives have been set to zero. Moreover, $\partial_t g_{tt}$ and $\partial_\varphi g_{tt}$ vanish due to the stationarity and axisymmetry of the problem. $\partial_\theta g_{tt}$ is also zero because the considered orbits are in the equatorial plane which is a surface of symmetry. Equation (\ref{e:geo}) then reduces to $g^{\mu r} \l[-\partial_r g_{tt}\r] = 2C\l(t\r) U^\mu$. Given the form of the metric (i.e. the use of quasi-isotropic coordinates), one can show that $g^{tr}$, $g^{\theta r}$ and $g^{\varphi r}$ all vanish. It first implies that one must have $C\l(t\r)=0$ which means that $t$ is indeed an affine parameter. Only the radial component of the geodesic equation is then not trivially satisfied and it reduces to 
\be
\partial_r g_{tt}=0.
\ee

\begin{figure}[!hbtp]
\includegraphics[width=12cm]{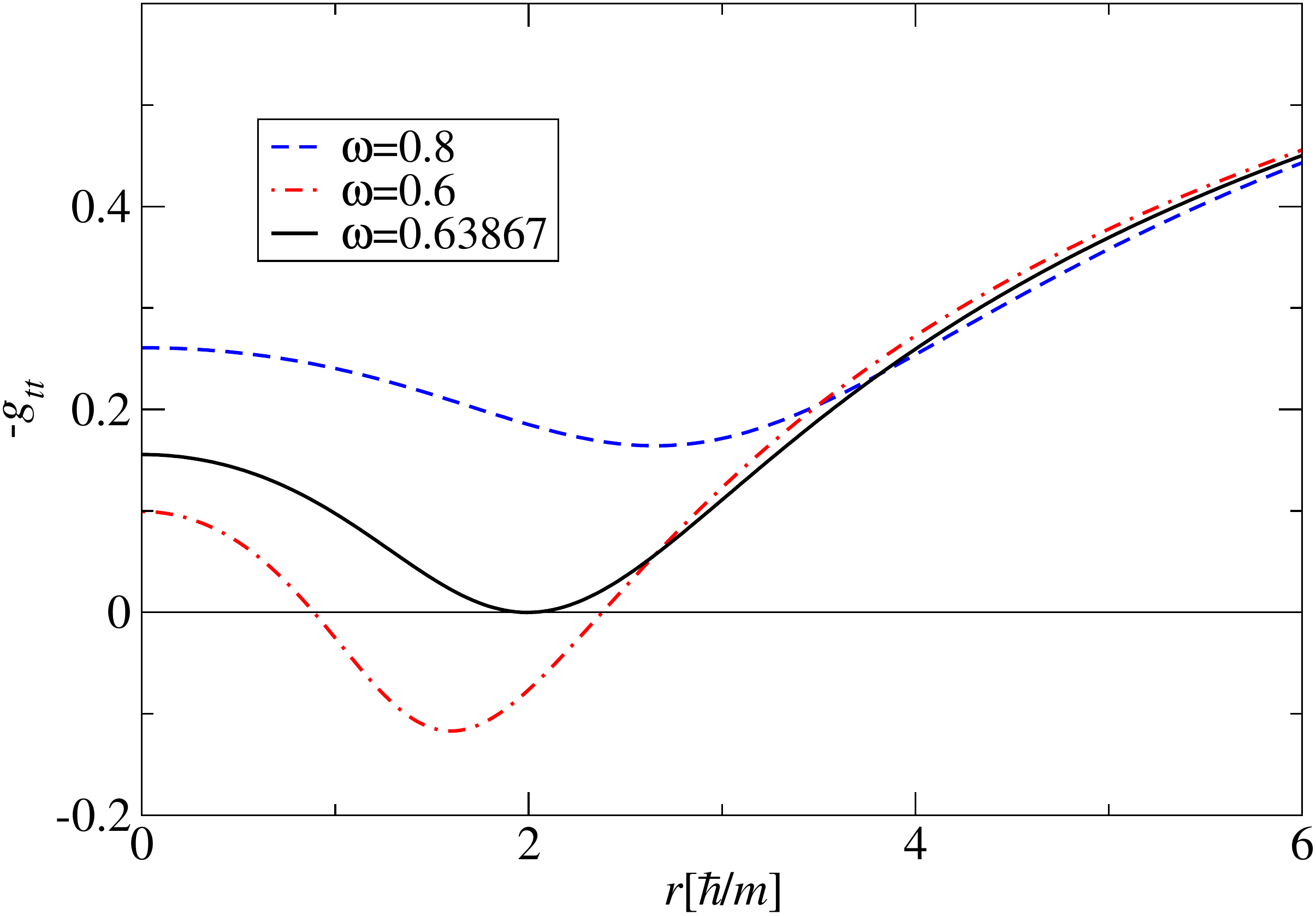}
\caption{\label{f:ergo} $-g_{tt}$ for three boson stars with $k=2$. The upper curve has no ergoregion. The lower one has a standard ergoregion in the shape of a torus. The middle curve corresponds to the critic case in which an ergoregion just starts to appear.}
\end{figure}

This condition on the derivative, along with the fact that $g_{tt}$ must vanish at the light points implies that they can only be situated exactly where an ergoregion starts to form. This is illustrated in Fig. \ref{f:ergo}, in which the quantity $-g_{tt}$ is plotted, in the plane $z=0$, for three different boson stars. The curve $\omega=0.8$ never goes to zero ; hence the associated boson star has no ergoregion. The curve $\omega=0.6$ gets below zero and vanishes at two different radii. They are the inner and outer radii of the ergoregion (for boson stars ergoregions have the shape of a torus ; see Sec. IV-E of Ref. \cite{GrandSG14}). The curve $\omega\approx 0.638$ corresponds to the case in which an ergoregion appears. $g_{tt}$ vanishes at just one point, which is also the minimum of the curve. This point is such that $g_{tt}=0$ and $\partial_r g_{tt}=0$ and so corresponds to a light point.

\begin{figure}[!hbtp]
\includegraphics[width=12cm]{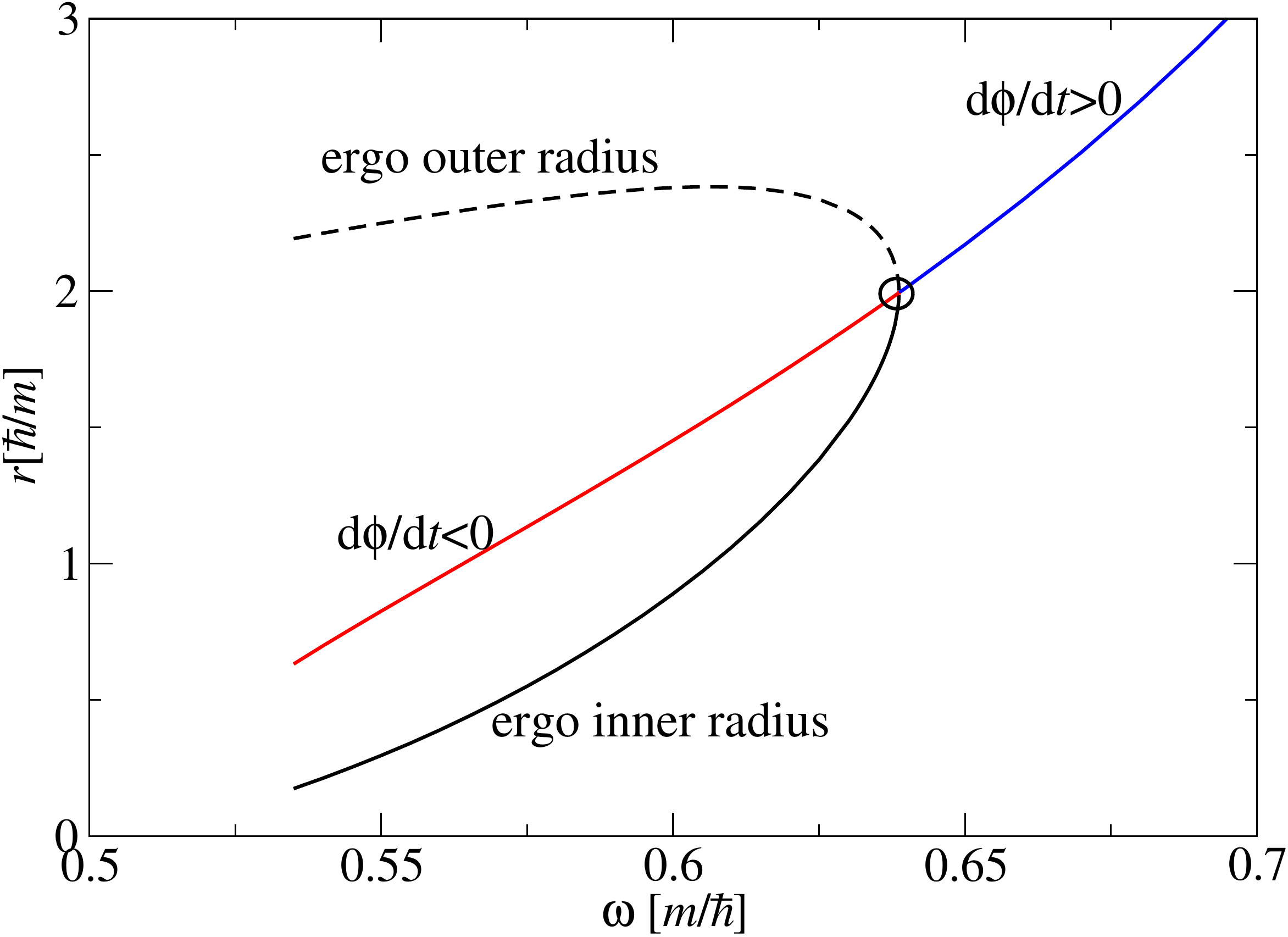}
\caption{\label{f:rad_points} For boson stars with $k=2$, inner and outer radii of the ergoregion (black curves). The red and blue curves denote the radius of the inner light ring, with different signs in the orbital frequency. All the curves intersect at the location of the circle which is the light point.}
\end{figure}

Figure \ref{f:rad_points} gives another illustration of the location of the light points (for the sequence $k=2$). The solid black curve denotes the inner  radius of the ergoregion and the dashed black one denotes the outer one. Those two curves join at $\omega\approx 0.638$ which is the frequency at which an ergoregion starts to develop. The blue curve shows the radius of the inner light rings for which ${\rm d}\varphi/{\rm d}t <0$ whereas the red curve denotes the same radius but for configurations with ${\rm d}\varphi/{\rm d}t >0$. It is clear from Fig. \ref{f:rad_points} that the light point lies at the intersection of the various curves, being the point where the ergoregion starts to develop. This is not so surprising. Ergoregions are defined as being locations where nothing can remain at rest, due to the frame-dragging effect generated by rotation. When it just appears only massless particles moving at the speed of light can overcome this effect and remain at rest. In other words, at the onset of the ergoregion, the frame-dragging effect is just compensated by the velocity of light. Each sequence of boson stars with $k>0$ admits a single light point. The values of $\omega$ and $R_{\rm minus}$ are summarized in Table \ref{t:points}.

It is worth discussing the existence of light points in the case of black holes. First, they cannot exist around rotating Kerr black holes. Indeed, as previously seen, light points exist only when the ergoregion is limited to a single ring. In the case of a Kerr black hole, it has a nonzero volume as soon as $a=0$, so as soon as one deviates from the Schwarzschild case. When $a=0$, photons at rest do indeed exist but they are on orbits exactly on the horizon (they are the null generators of the event horizon). However, and this is the main difference from the boson star case, they correspond to unstable orbits. Stable light points are therefore very specific to boson star spacetimes.

\begin{table}
\caption[]{\label{t:points}
Parameters of the light points for $k=1$ to $k=4$.
} 
\begin{tabular}{| c | c |  c | }
  \hline
  $k$ & $\omega$ & $R_{\rm minus}$  \\
\hline
$1$ & $0.6582$ & $0.779$ \\
$2$ & $0.6387$ & $1.99$ \\
$3$ & $0.6383$ & $3.36$ \\
$4$ & $0.6401$ & $4.83$ \\
\hline
  \end{tabular}
\end{table}

\begin{figure}[!hbtp]
\includegraphics[width=8cm]{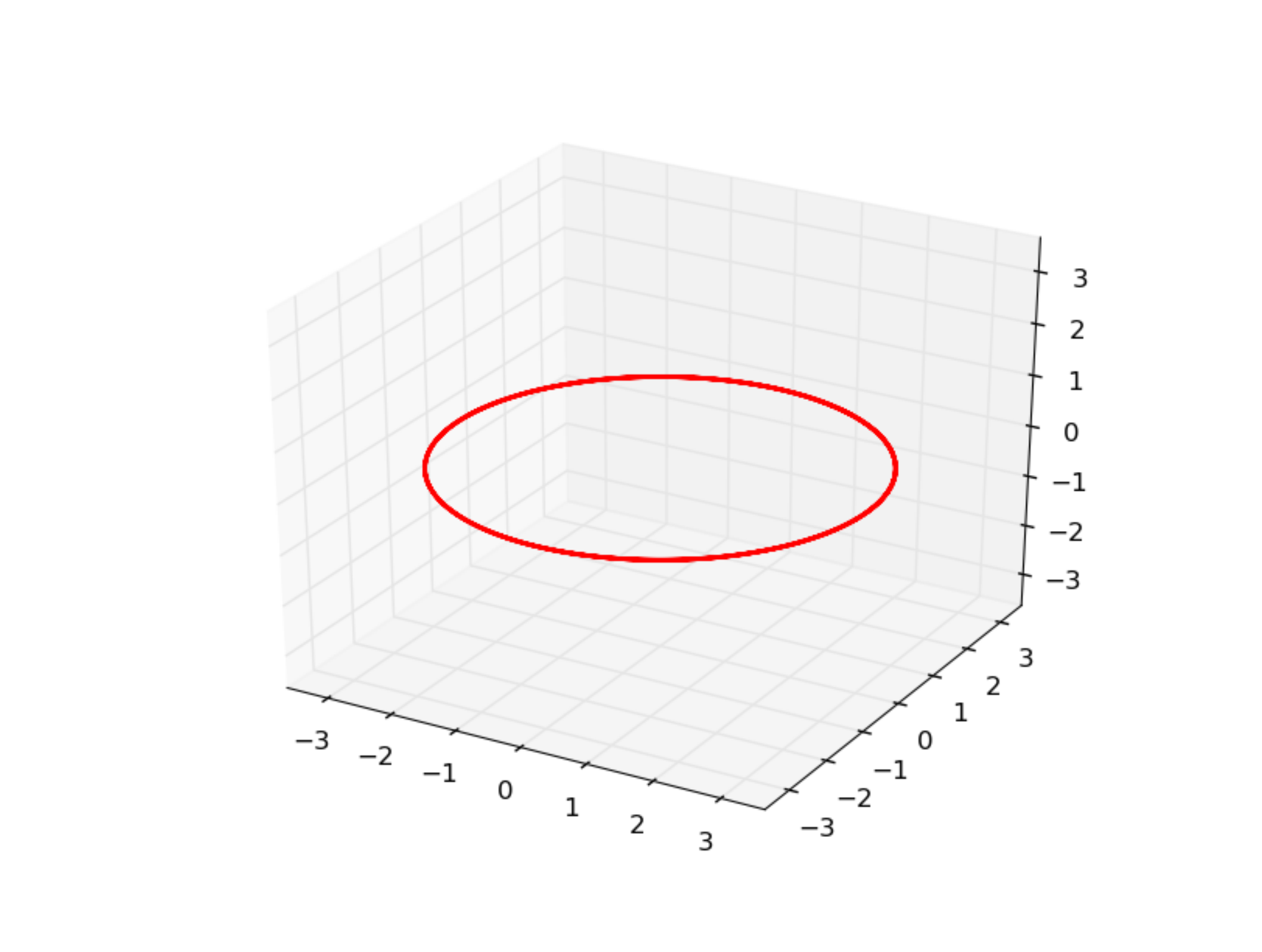}
\includegraphics[width=8cm]{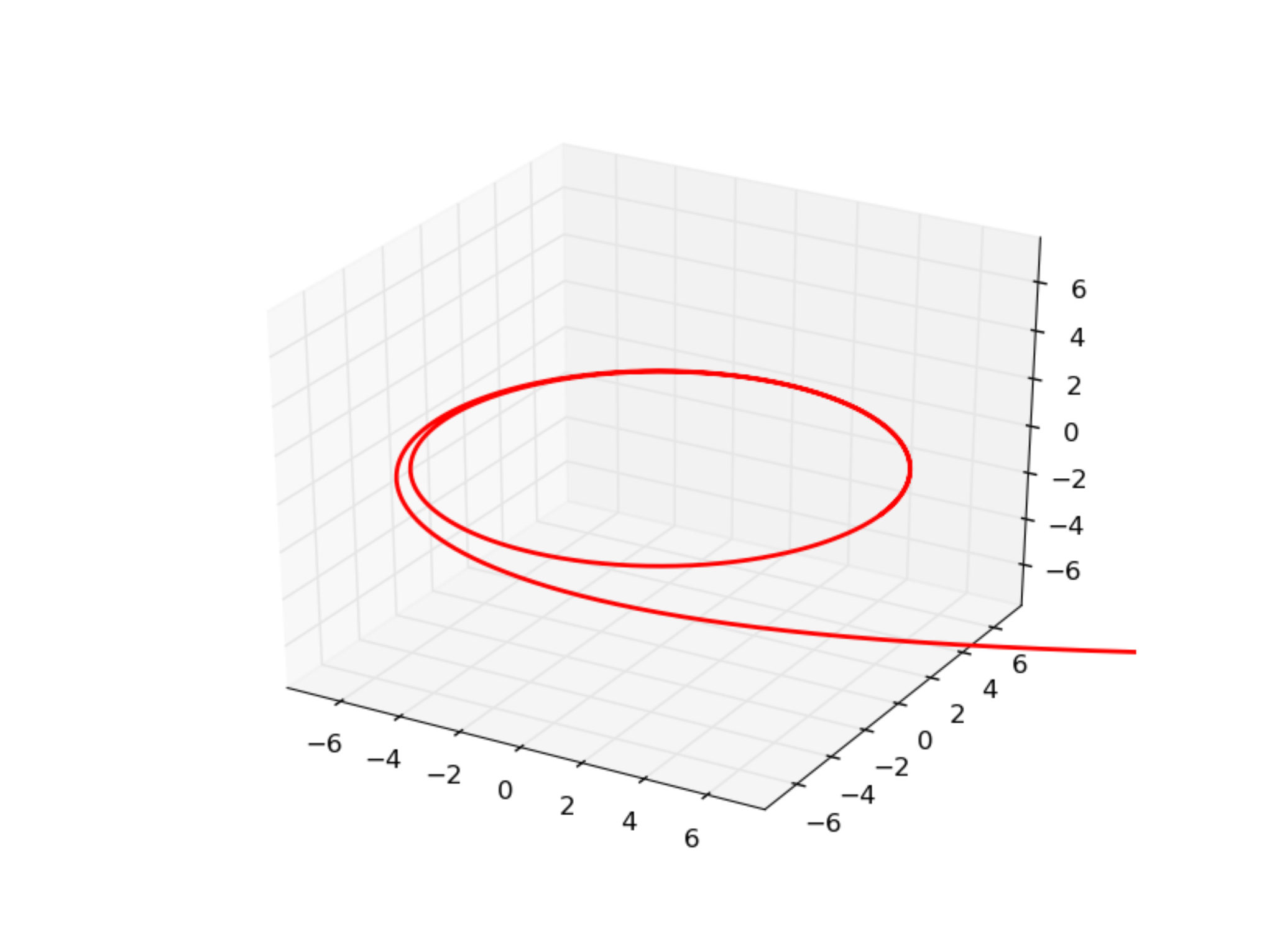}
\includegraphics[width=8cm]{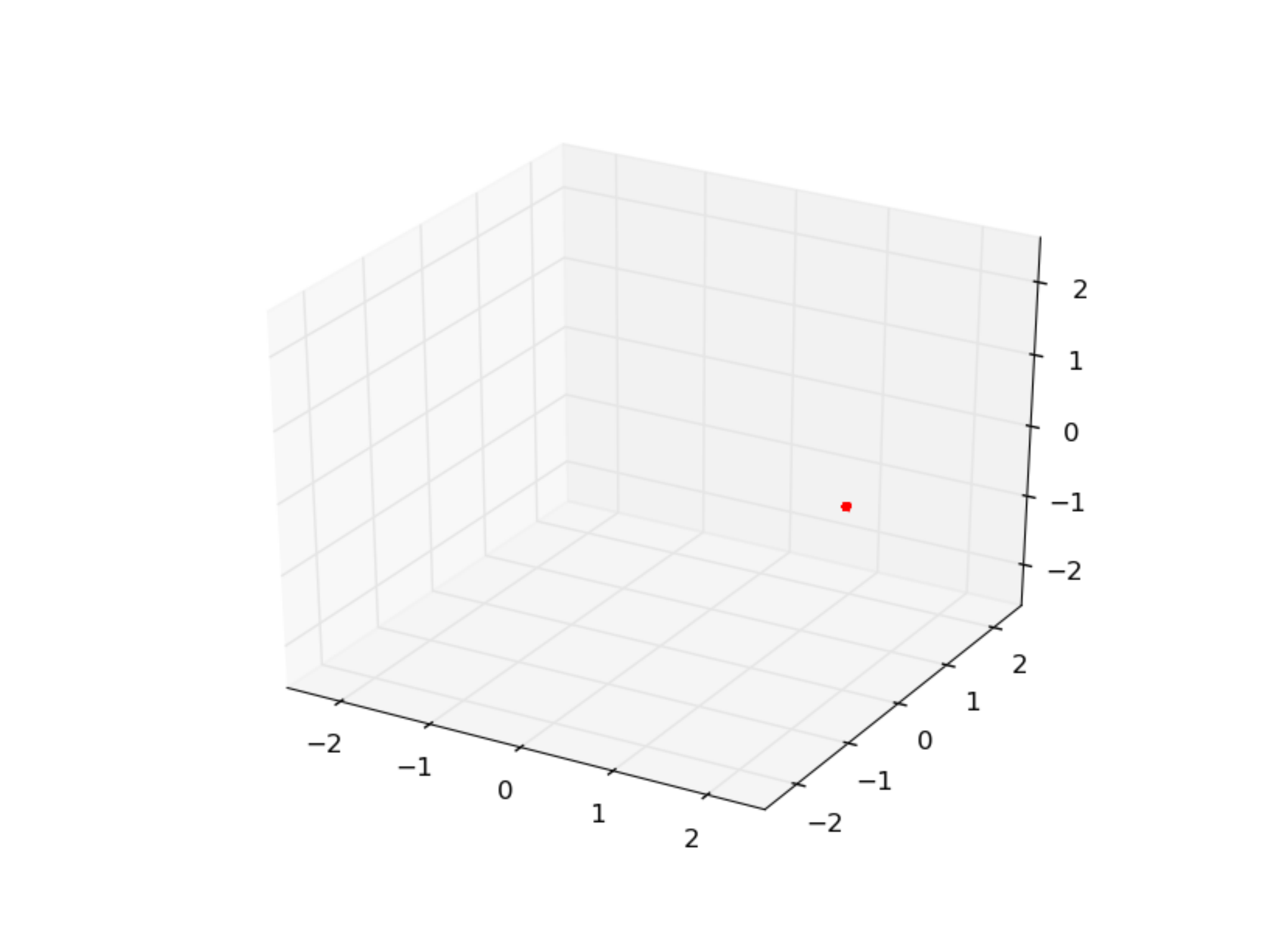}
\caption{\label{f:gyoto_orbits} Orbits integrated numerically. The first panel corresponds to a stable inner light ring whereas the second one depicts the orbit of an unstable outer one (for $k=2$ and $\omega=0.7$). The last panel shows the light point of the $k=2$ sequence.}
\end{figure}

The results can be checked by integrating numerically the geodesic equation. This is done using the tool Gyoto \cite{VincePGP11}. One can compute the radius of a light ring, put a photon there, with the right angular frequency [i.e. given by Eq. (\ref{e:dpdt})] ; and check whether the computed orbit is indeed circular. Figure \ref{f:gyoto_orbits} shows such computations, in the case $k=2$. The first panel corresponds to the inner light ring ($R_{\rm minus} = 3.11139$ and ${\rm d}\varphi/{\rm d}t = -0.0404863$) and the second panel corresponds the outer light ring ($R_{\rm plus} = 7.07489$ and ${\rm d}\varphi/{\rm d}t = -0.0637682$), both for $\omega=0.7$. The photon on the inner light ring remains nicely on a circular orbit. The one on the outer light ring starts by describing a circular orbit but is eventually kicked out. This is consistent with the fact that only the inner light ring is stable, corresponding to a minimum of the effective potential. The last panel shows the light point of the $k=2$ sequence (i.e. $\omega=0.6387$ and $R=1.99$). As expected, the photon stays at the same spatial location and so appears as a single point in Fig. \ref{f:gyoto_orbits}.

\section{Conclusion}\label{s:ccl}

In this paper, models of rotating boson stars are studied. They were obtained numerically in a previous work \cite{GrandSG14}. It is confirmed that rotating boson stars are objects that can reach very high compactness. This implies that boson stars can exhibit effects in which strong gravity is crucial such as the existence of closed orbits of photons. The existence of those orbits, called light rings, is confirmed for a large class of boson stars. Light rings could have strong implications on the stability of the boson stars themselves \cite{CardoCMOP14}.

A new class of light rings, which corresponds to stable trajectories in which the photons are at rest, with respect to an observer at infinity, is exhibited. It is proposed to call those orbits light points. Their existence is closely linked to the appearance of an ergoregion. They are very specific of boson stars in the sense that they cannot exist around black holes. Indeed in that case they are located exactly on the horizon and correspond to unstable trajectories. It is believed that this is the first time that those stable light points are discussed. Their existence has been confirmed by a direct integration of the geodesic equation.

\acknowledgments{The author thanks V. Cardoso and E. Gourgoulhon for convincing him to study light rings around boson stars.}

\end{document}